\documentclass[fleqn,12pt,twoside]{article}
\usepackage{espcrc1}
\usepackage{epsfig}

\usepackage{graphicx}
\usepackage[figuresright]{rotating}

\newcommand{\AmS}{{\protect\the\textfont2
  A\kern-.1667em\lower.5ex\hbox{M}\kern-.125emS}}

\title{Symmetry patterns in the $(N,\Delta )$ spectrum}

\author{P. Gonz\'alez, J. Vijande \address{Dpto. de F\'\i sica Te\'orica 
        and IFIC, Universidad de Valencia - CSIC, \\
        E-46100 Burjasot, Valencia, Spain},
        A. Valcarce\address{Grupo de F\'\i sica Nuclear and IUFFyM,
        Universidad de Salamanca,\\
        E-37008 Salamanca, Spain},
        and
        H. Garcilazo\address{Escuela Superior de F\'\i sica y Matem\'aticas,
        Instituto Polit\'ecnico Nacional, \\
        Edificio 9, 07738 M\'exico D.F., M\'exico}}
       
\begin{document}

\maketitle

\begin{abstract}
We revise the role played by symmetry in the study of \ the low-lying baryon
spectrum and comment on the difficulties when trying to generalize the
symmetry pattern to higher energy states. We show that for the $(N,\Delta )$
part such a generalization is plausible allowing the identification of
spectral regularities and the prediction of until now non-identified
resonances.
\end{abstract}

\section{Introduction}

The PDG Baryon Summary Table \cite{Eid04} contains 123 resonances. This
richness is telling us about the existence, properties and dynamics of the
intrabaryon constituents. In order to extract this physical content, the
knowledge of spectral patterns is of great help. For instance the
classification of the low-lying baryons according to $SU(3)_{\rm flavor}$
multiplets in Gell Mann's {\it eightfold way} revealed the existence of quarks
and made clear spectral regularities from which to predict new states as the 
$\Omega $ particle. The consideration of additional spin and orbital degrees
of freedom demanded the enlargement of the symmetry group. The assumption
that quarks feel a rotationally invariant potential\ resulted in a 
$SU(6)\otimes O(3)$ pattern. Mass differences inside the $(N,L^{P})$
multiplets ($N$ standing for the $SU(6)$ multiplet) pointed out the need to
implement a symmetry breaking in the dynamics. The inclusion of a one gluon
exchange chromomagnetic quark-quark interaction allowed for a correct
description of the observed mass splitting \cite{Clo79}.

\begin{figure}[t]
\vspace*{-2cm}
\hspace{-3cm}
\mbox{\epsfig{file=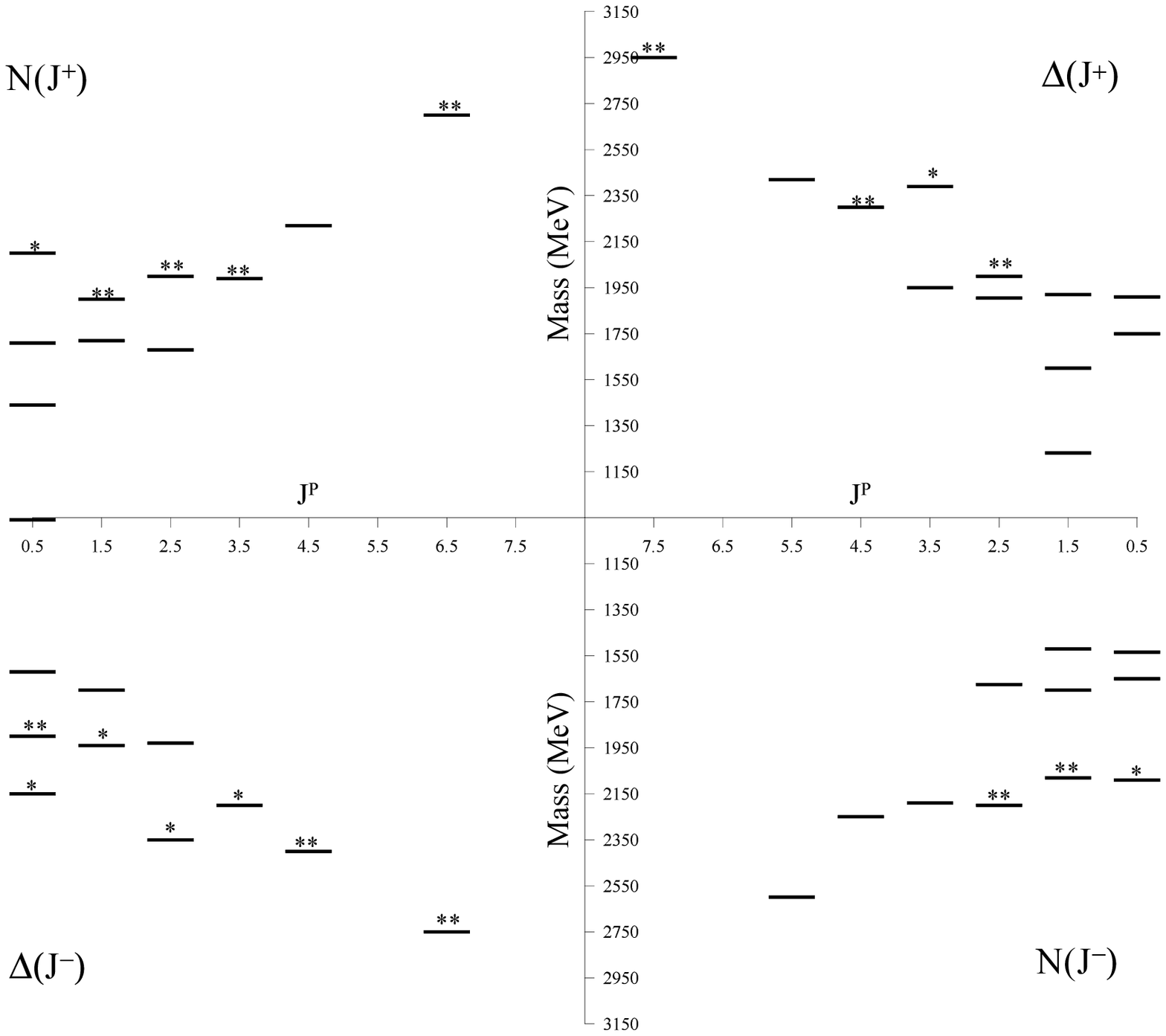,angle=-90,width=200mm}}
\vspace*{-2.2cm}
\caption{Nucleon and $\Delta$ spectra from PDG \protect\cite{Eid04}. 
Stars have been omitted for four and three star resonances.}
\label{fig1}
\end{figure}

When going to higher energy states the ascription of resonances to
multiplets becomes much more difficult because of the different spin-orbital
structures entering as resonance components. Furthermore the same validity
of $SU(6)\otimes O(3)$ as a symmetry group may be under suspicion if
relativistic effects, mixing orbital and spin degrees of freedom, becomes
relevant. An unambiguous baryon quantum number assignment demands in
practice two conditions to be satisfied: first the use of a complete data
set and second the use of a dynamical model being able to reproduce the
number and ordering of known resonances. These conditions can be rather
well satisfied for the lightest-quark $(u,d)$-baryon spectrum to which we
shall restrict hereforth. This $(N, \Delta )$ spectrum, containing 45
known resonances, 25 of them well established experimentally, is represented
in Fig. \ref{fig1}.

\begin{table}[htb]
\caption{Positive parity $N$ and $\Delta $ states classified in multiplets
of $SU(4) \otimes O(3)$ (up to $\simeq 3000$ MeV mass).
Experimental data are from PDG \protect\cite{Eid04}. Stars have been omitted
for four-star resonances. States denoted by a question mark correspond to
predicted resonances, Fig. \protect\ref{fig2}, that do not appear in the PDG.}
\label{t1}
\newcommand{\m}{\hphantom{$-$}}
\newcommand{\cc}[1]{\multicolumn{1}{c}{#1}}
\renewcommand{\tabcolsep}{2pc} 
\renewcommand{\arraystretch}{1.2} 
\begin{tabular}{@{}|c|c|c|} \hline
$(N,L^{P})$ & $S=1/2$ & $S=3/2$ \\ \hline
$(20_{S},0^{+})$ & $N_{S}(1/2^{+})(940)$ & $\Delta (3/2^{+})(1232)$ \\ \hline
$(20_{S},2^{+})$ & $N_{S}(5/2^{+})$ & $\Delta (7/2^{+})(1950)$ \\ \hline
$(20_{S},4^{+})$ & $N_{S}(9/2^{+})$ & $\Delta (11/2^{+})(2420)$ \\ \hline
$(20_{S},6^{+})$ & $N_{S}(13/2^{+})$ & $\Delta (15/2^{+})(2950)(\ast \ast )$
\\ \hline\hline
$(20_{M},0^{+})$ & $N_{M}(1/2^{+})(1710),\Delta (1/2^{+})(1750)$ & $%
N_{M}(3/2^{+})$ \\ \hline
$(20_{M},2^{+})$ & $N_{M}(5/2^{+}),\Delta (5/2^{+})(1905)$ & $%
N(7/2^{+})(1990)(\ast \ast )$ \\ \hline
$(20_{M},4^{+})$ & $N_{M}(9/2^{+}),\Delta (9/2^{+})(2300)(\ast \ast )$ & $%
N(11/2^{+})(?)$ \\ \hline
$(20_{M},6^{+})$ & $N_{M}(13/2^{+}), \Delta (13/2^+)(?)$ & $N(15/2^{+})(?)$ \\ \hline
\end{tabular}
\end{table}

\begin{figure}[htb]
\vspace*{-2cm}
\hspace{-2cm}
\mbox{\epsfig{file=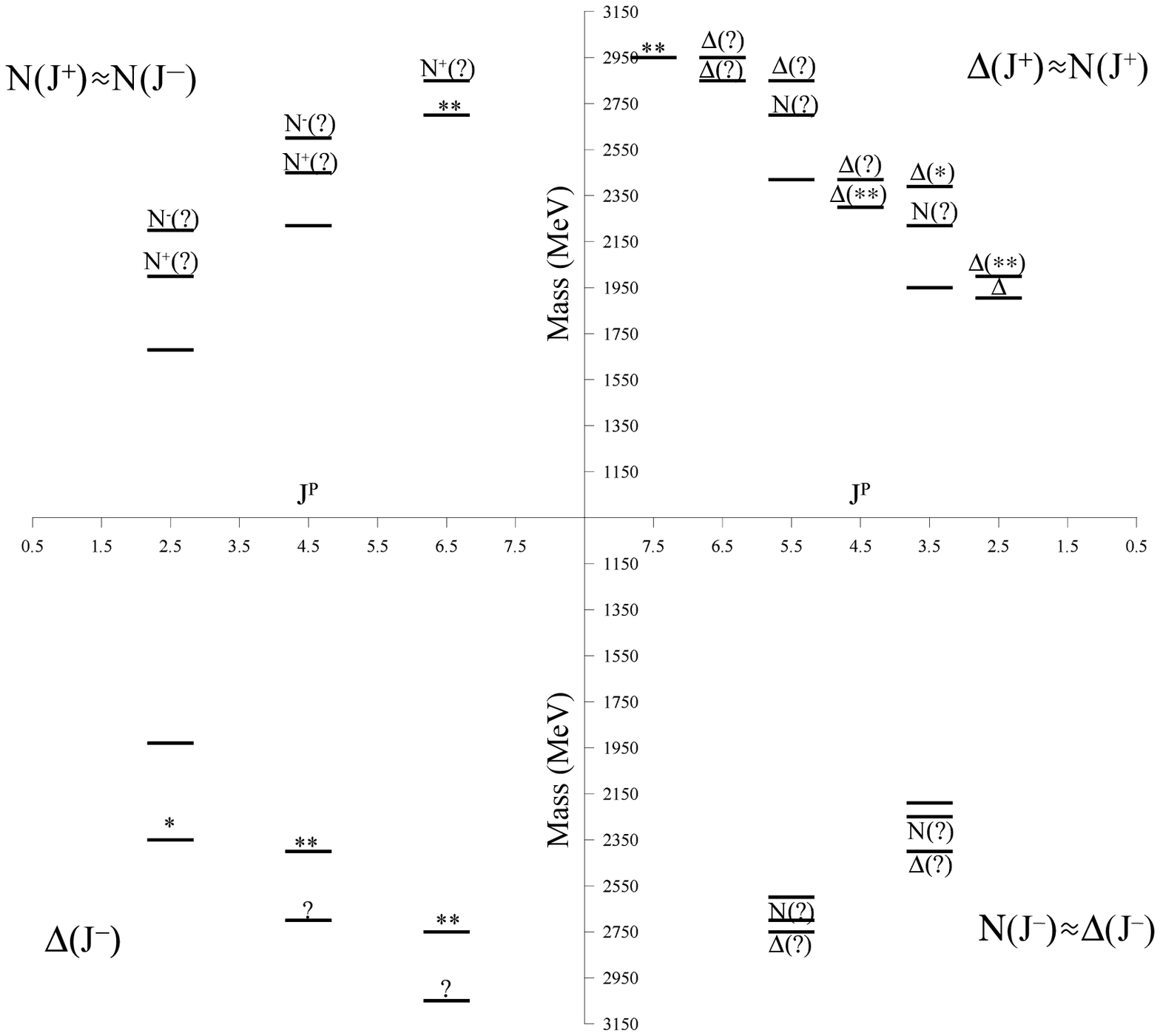,angle=-90,width=200mm}}
\vspace*{-2.2cm}
\caption{Nucleon and $\Delta$ spectrum (data and predictions)
for $J \geq 5/2$. Predicted states are indicated by a question mark. 
Those $\Delta$'s ground state not degenerate with $N$'s as 
well as first non-radial excited states for $\Delta$'s and $N$'s 
are explicitly indicated by a  corresponding $\Delta$ or $N$ symbol.}
\label{fig2}
\end{figure}

\begin{table}[htb]
\caption{Same as Table \protect\ref{t1} for negative parity $N$ 
and $\Delta $ states.}
\label{t2}
\newcommand{\m}{\hphantom{$-$}}
\newcommand{\cc}[1]{\multicolumn{1}{c}{#1}}
\renewcommand{\tabcolsep}{2pc} 
\renewcommand{\arraystretch}{1.2} 
\begin{tabular}{@{}|c|c|c|} \hline
$(N,L^{P})$ & $S=1/2$ & $S=3/2$ \\ \hline
$(20_{M},1^{-})$ & $N_{M}(3/2^{-})(1520),\Delta (3/2^{-})(1700)$ & $%
N(5/2^{-})(1675)$ \\ \hline
$(20_{M},3^{-})$ & $N_{M}(7/2^{-}),\Delta (7/2^{-})(2200)(\ast )$ & $%
N(9/2^{-})(2250)$ \\ \hline
$(20_{M},5^{-})$ & $N_{M}(11/2^{-}),\Delta (11/2^{-})(?)$ & $N(13/2^{-})(?)$
\\ \hline\hline
$(20_{S},1^{-})$ & $N_{S}(3/2^{-})$ & $\Delta (5/2^{-})(1930)(\ast \ast \ast
)$ \\ \hline
$(20_{S},3^{-})$ & $N_{S}(7/2^{-})$ & $\Delta (9/2^{-})(2400)(\ast \ast )$
\\ \hline
$(20_{S},5^{-})$ & $N_{S}(11/2^{-})$ & $\Delta (13/2^{-})(2750)(\ast \ast )$\\ \hline
\end{tabular}
\end{table}

\section{Dynamical model and the symmetry pattern}

A dynamical model satisfying the second condition above can be built from a
minimal quark potential model (containing a linear confinement plus a
hiperfine one gluon exchange interaction) by incorporating
screening as an effective mechanism to take, at least partially into
account, the effect derived from the opening of decay channels. 
Screening can be put in the form of a
saturating distance beyond which the quark-quark potential becomes a
constant \cite{Gon06}. Except for the spin-spin
piece the model is approximately $SU(4)(\supset SU(2)_{\rm spin}\otimes
SU(2)_{\rm isospin})\otimes O(3)$ symmetric. Hence we expect the baryons to be
classified according to $SU(4)$ 20plets (since $4\otimes 4\otimes
4=20_{S}\oplus 20_{M} \oplus 20_{M} \oplus \overline{4})$ with defined orbital angular momentum 
$L$ and parity $P$. This is confirmed by the analysis of the dominant
configurations for the ground and first non-radial excited states of any 
$J^{P}$resonance up to a mass of 3 GeV \cite{Gon06}. The multiplet
pattern appears in Tables \ref{t1} and \ref{t2}. 

The subindexes $S$ and $M$ indicate
multiplet states that mix to give rise to experimental resonances
(nonetheless when one of the configurations ($S$ or $M)$ is clearly dominant
we have assigned to it the experimental mass: 
$N_{S}(1/2^{+})(940),N_{M}(1/2^{+})(1710),...)$. Thus for example 
$N(5/2^{+})(1680)$ gets in our model a 62\% of $N_{S}(5/2^{+})$ and a 34\%
of $N_{M}(5/2^{+})$ whereas $N(5/2^{+})(2000)$ gets 35\% of 
$N_{S}(5/2^{+}) $ and 64\% of $N_{M}(5/2^{+})$. A comparative analysis
between the model masses and data might even indicate less mixing than
calculated by the model, what would make the symmetry scheme more
predictive. This seems to be confirmed by the spectral regularities observed
for $J\geq 5/2$ when one identifies multiplet states with known
resonances, for instance $N_{S}(5/2^{+})\approx N(5/2^{+})(1680)$,
$N_{M}(5/2^{+})\approx N(5/2^{+})(2000)$ and so on (one exception is 
$N(3/2^{+})(1720)$ with the same dominant configuration than $N(5/2^{+})(1680)$
due to the spin-spin interaction). These spectral regularities can be
summarized as:

i) Intermultiplet energy difference: $E_{N,\Delta }(J+2)-E_{N,\Delta
}(J)\approx 400-500$ MeV.

ii) $N-\Delta $ ground state degeneracies: $N(J^{\pm})\approx \Delta (J^{\pm})$ for 
$J= \frac{4n+3}{2}$, $n=1,2...$.

iii) $N$ ground state parity doublets: $N(J^{+})\approx N(J^{-})$ for $J=\frac{4n+1}{2}$, 
$n=1,2...$.

iv) First non-radial excitations: $(N(J),\Delta (J))^{^{\bullet }}\approx
(N(J+1),\Delta (J+1))$.

\noindent
The extension of this pattern up to 3 GeV drives to the prediction of
until now undetected resonances as shown in Fig. \ref{fig2}, 
containing the $\left[ J=5/2 - J=15/2\right] $ 
ground and first non-radial excitations.
It is worth to mention that the existence of nucleon parity doublets
is a consequence of the almost exact cancellation of a bigger 
repulsion (due to a bigger $K$ and $L$) and a bigger attraction (due 
to $S=1/2$) for $N^+$ states. No parallel degeneracy for $\Delta$'s
is found against theoretical proposals \cite{GloXX} that have
been questioned recently \cite{Jaf05}. Though
new experimental data, to confirm or discard the validity of our predictions,
are definitely needed we think the {\it symmetry way} can be of great 
interest to guide experimental searches.

\bigskip

This work has been partially funded by MCyT
under Contract No. FPA2004-05616, by JCyL under
Contract No. SA104/04, and by GV under Contract No.
GV05/276.


\begin{thebibliography}{9}
\bibitem{Eid04} S. Eidelman {\it et al.}, Phys. Lett. B 592 (2004) 1.
\bibitem{Clo79} F.E. Close, An Introduction to Quarks and Partons,
Academic Press, 1979.
\bibitem{Gon06} P. Gonz\'alez, J. Vijande, A.Valcarce, and H. Garcilazo, 
Eur. Phys. J. A 29 (2006) 235.
\bibitem{GloXX} T.D. Cohen and L. Ya. Glozman, Phys. Rev. D 65
(2002) 016006.
\bibitem{Jaf05} R.L. Jaffe, D. Pirjol, and A. Scardicchio,
Phys. Rev. Lett. 96 (2006) 121601.
\end{thebibliography}
\end{document}